\documentclass[aps,prl,reprint,graphicx]{revtex4-1}
\usepackage{amsmath}
\usepackage{amssymb}
\usepackage{graphicx}  
\begin{document}

\title{3D printing of gas jet nozzles for laser-plasma accelerators}

\author{A. D\"opp}
\author{E. Guillaume}
\author{C. Thaury}
\author{J. Gautier}
\author{K. Ta Phuoc}
\author{V. Malka}

\address{LOA, ENSTA ParisTech - CNRS - \'Ecole Polytechnique - Universit\'e Paris-Saclay, 828 Boulevard des Mar\'echaux, 91762 Palaiseau Cedex, France}

\begin{abstract}

Recent results on laser wakefield acceleration in tailored plasma channels have underlined the importance of controlling the density profile of the gas target. In particular it was reported that appropriate density tailoring can result in improved injection, acceleration and collimation of laser-accelerated electron beams. To achieve such profiles innovative target designs are required. For this purpose we have reviewed the usage of additive layer manufacturing, commonly known as 3D printing, in order to produce gas jet nozzles. Notably we have compared the performance of two industry standard techniques, namely selective laser sintering (SLS) and stereolithography (SLA). Furthermore we have used the common fused deposition modeling (FDM) to reproduce basic gas jet designs and used SLA and SLS for more sophisticated nozzle designs. The nozzles are characterized interferometrically and used for electron acceleration experiments with the \textsc{Salle Jaune} terawatt laser at Laboratoire d'Optique Appliqu\'ee. 

\end{abstract}

\maketitle

Particle accelerators are an essential tool in science, industry and medicine. While a century of R\&D has lead to a high level of control and stability, field emission and subsequent vacuum breakdown still limit the maximum field gradients to around 100 MV/m \cite{Wiedemann:2015ws}. This bottleneck prevents high energy accelerators from becoming more compact and affordable. Plasma-based accelerators overcome these limitations by use of a pre-ionized medium and can thus reach higher acceleration gradients, in excess of 100 GV/m \cite{Esarey:2009ks}. In particular it was observed that electrons can be accelerated to highly relativistic energies in the wake of an intense laser pulse propagating through a plasma \cite{Malka:2002eu}. Moreover, it has been demonstrated that the kinetic energy of these electron beams can be converted on a millimeter scale into energetic photon beams using e.g. bremsstrahlung conversion \cite{Glinec:2005ve,Dopp:2016tv}, magnetic undulators \cite{Schlenvoigt:2008bg}, inverse Compton backscattering \cite{TaPhuoc:2012cg,Powers:2013bx} or betatron emission from plasma wigglers \cite{Rousse:2004tc}.

The density profile of the plasma target plays a crucial role for the operation of a laser-wakefield accelerator. In particular, many recent innovations were achieved by means of target engineering. For instance it has been shown that longitudinal density tailoring can be used to localize electron injection \cite{Schmid:2010ih,Buck:2013gs}, increase the beam energy \cite{Dopp:2016gm}, reduce the beam energy spread \cite{Guillaume:2015dia} and the beam divergence \cite{Thaury:2015cg}. Target engineering is therefore very important for the advance of the research field. However, there are a number of problems with the current target manufacturing technology. One is that the targets become more and more complex and traditional manufacturing techniques are brought to their limits. Also, most high-intensity lasers are located at user facilities and laser-plasma acceleration experiments have a typical duration of a few weeks. With conventional production chains it is therefore often impossible to innovate targets during a campaign. As an alternative we have investigated the usage of 3D printers for gas jet manufacturing. \\

The paper is structured as follows: First, we give an overview of gas targets for laser wakefield acceleration and the most common 3D printing technologies. Then we present results using different techniques and compare their performance. Last, we discuss how 3D printed nozzles perform in experiments on laser-driven electron acceleration and what are the merits of the technology in this special field of applications.

\section{Gas jet targets for laser-plasma accelerators}

Laser wakefield accelerators rely on a plasma to act as both electron injector and accelerator \cite{Esarey:2009ks}. This plasma is created via ionization of a gas target and in order to permit a laser pulse to propagate, the electron density $n_e$ has to be below the critical density $n_c\simeq 1.7\times 10^{21}\times (\lambda_0[\mu m])^{-2}$. While propagation through the plasma, the ponderomotive force of the laser pulse excites a plasma wave in its wake that serves as accelerating structure. However, as the laser propagates in an underdense plasma at a group velocity $v_g\simeq (1-n_e/2n_c)c_0$, the laser driver and the electrons (with a velocity close to the speed of light in vacuum $v_e\simeq c_0$) slowly move with respect to each other. Once the electrons have reached a decelerating region of the plasma wave they are considered dephased and will start loosing energy. In a flat density profile the resulting dephasing length $L_d \propto n_e^{-3/2}$ then defines the maximum attainable beam energy, leading to an approximate scaling $E_{max}\propto n_e^{-1}$, see Ref \cite{Esarey:2009ks}. As operation above the dephasing length is inefficient, the accelerator length is usually optimized around the dephasing length. This relation is illustrated in Fig.1, which shows a collection of experimental data from \cite{Mangles:2016wp}, with the scalings of beam energy and matched plasma density. While longer, low-density targets are adequate to produce beams of highest energy, high density targets can be of interest to produce beams with high charge. Here the reduced group velocity of the laser can facilitate self-injection of electrons into the accelerator \cite{Benedetti:2013fy}. Furthermore, the accelerator length is limited by the laser power: Terawatt-class facilities often use jets of a few millimeter length, while petawatt-class lasers employ centimeter-scale targets.

\begin{figure}[t]
\centering
\includegraphics[width=1.0\linewidth]{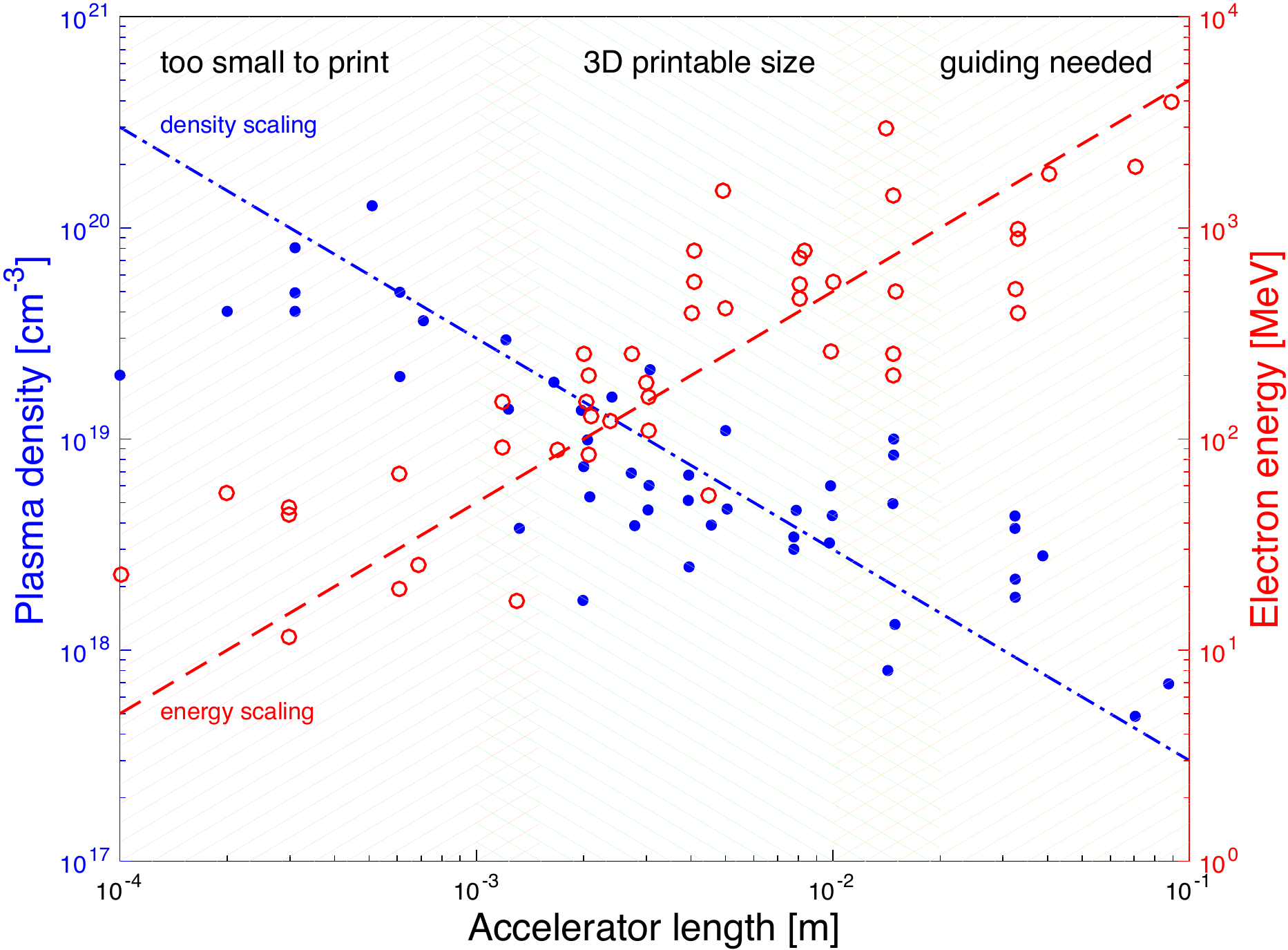}
\caption{Scalings of dephasing-limited laser-plasma accelerators. As discussed in \cite{Esarey:2009ks}, higher beam energies can be reached at reduced plasma density, which in turn requires longer acceleration length due to the reduced wakefield amplitude. The lines shows the basic scalings, which agree well with the marked data from various experiments on laser wakefield acceleration \cite{Mangles:2016wp}. As we will discuss in this paper, structures smaller than a millimeter can currently not be produced using common 3D printers. The technology is ideal for a range between 2 and 10 millimeter nozzle size. Above this capillary targets may be the better choice due to their additional guiding capabilities.}
\end{figure}

Several different types of such targets have been developed and the most common designs are discharge and dielectric capillaries, gas cells and gas jets. The first three types have a similar layout, where the laser is focus into a tube with a diameter in the order of hundred microns. Note that careful alignment of the laser beam is imperative since even a slight misalignment can damage or destroy the target.
Discharge targets provide a preformed plasma with a transverse density gradient that can provide auxiliary guiding of the laser pulse \cite{Butler:2002ho}. The discharge-based accelerator can therefore be operated at lower plasma density than self-guided laser-plasma accelerators and is often chosen for high energy experiments \cite{Leemans:2006ux,Leemans:2014kp}. However, in the past the complexity of the discharge target has lead to stability issues. Dielectric capillary tubes work as waveguides and therefore provide additional guiding of outer Airy laser modes \cite{Ju:2012jz}. Gas cells have a similar design, but provide no additional guiding capabilities. Yet they offer certain advantages, e.g. some designs allow to easily vary the acceleration length \cite{Corde:2013gj}. In both cases the gas is supplied from a reservoir of some hundred millibar backing pressure and the gas density is constant inside the target, with entrance and exit gradients of the order of their diameter \cite{Ju:2012es}.

Gas jets are different from the above layouts as the laser in focused onto a gas that flows freely into the vacuum chamber. A gas jet primarily consists of a high-pressure gas valve and an individually designed exit nozzle. Many laboratories rely on Series 9 pulse vales by Parker Hannifin Corp., which operate at up to 50-80 bar and a sub-millisecond reaction time. Onto the valve exit different nozzles can be mounted. These nozzles are produced usually using computerized numerical control (CNC) milling in aluminum, with most designs relying on a conical De Laval layout that leads to a supersonic gas flow downstream \cite{Semushin:2001fl,Schmid:2012ki}. The nozzles have usually exit diameters of a few millimeters, but also nozzles of more than a centimeter diameter have been used for GeV electron acceleration \cite{Najmudin:2014uu}. A very particular type gas jet nozzles has been developed for laser-driven proton acceleration. Here near-critical plasma densities are reached, which requires operation at very high pressure ($>$100 bar) and small exit diameters ($<$1 mm) \cite{Sylla:2012hl}. 

There are a number of reasons why gas jets are the most common type of gas target used for laser wakefield acceleration.
First, their open geometry makes them much simpler to align and offers good accessibility for diagnostics, which is interesting for prototyping and proof-of-concept setups. But furthermore their superior stability and durability makes them a frequent choice for permanent setups. Last, the longitudinal density profile in gas jets can be tailored to an extend that has not yet been demonstrated with other targets. For example, the gas flow from a single nozzle can be manipulated with a blade to create sharp transitions \cite{Schmid:2010ih,Guillaume:2015dia} or the flow of multiple exit nozzles can be combined to create up or downramps in the profile \cite{Golovin:2015hc}.

\section{Additive manufacturing technologies (3D printing)}

\begin{figure*}[t]
\centering
\includegraphics[width=1.0\linewidth]{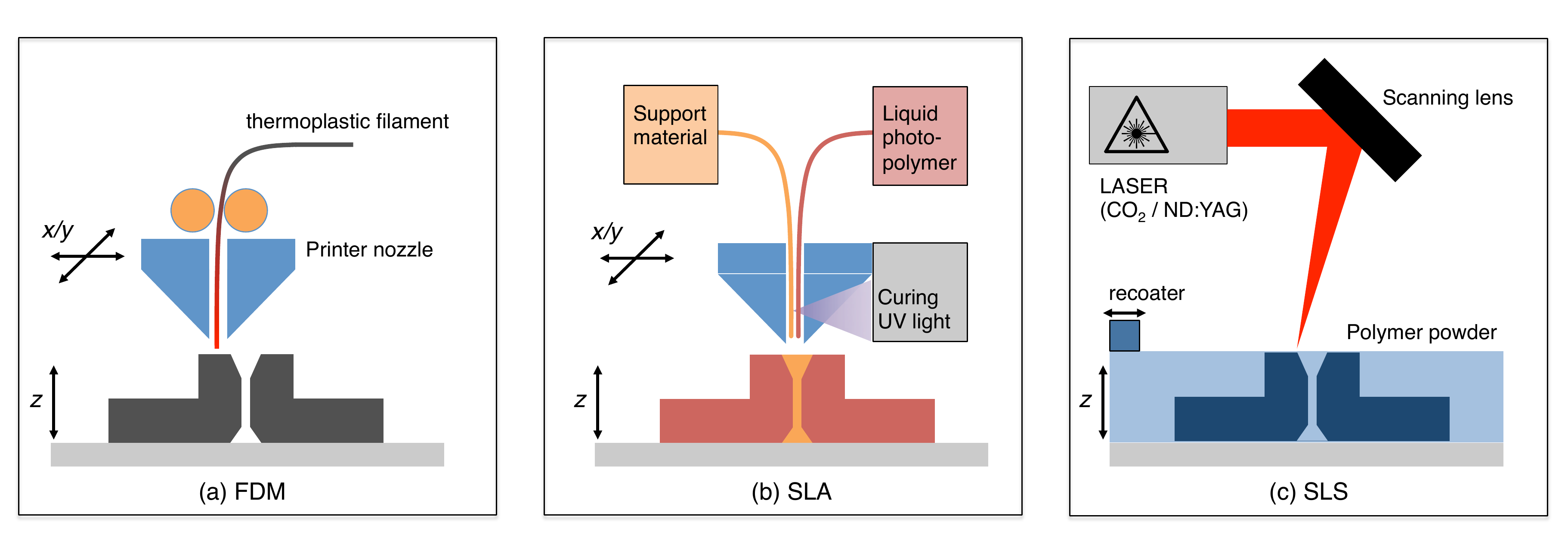}
\caption{Illustration of three different additive manufacturing techniques. (a) Fused deposition modeling, (b) PolyJet, a variation of stereolithography and (c) selective laser sintering.}
\end{figure*}

What is commonly called 3D printing \cite{Sachs:1992cb} is a number of different technologies for additive layer manufacturing \cite{Levy:2003ie,Wong:2012iq,Conner:2014hr}. These technologies have been essentially developed since the 1980s and have recently drawn a lot of public attention \cite{Berman:2014ko}. Also many original patents recently expired, which may further support the growth of this new industry \cite{Hornick:2013wx}. While many applications e.g. in engineering, biotechnology and chemistry have been considered \cite{Gross:2014jxa}, the technology still faces many challenges and each novel application has to be carefully reviewed \cite{Gao:2015bs}.

There are a number of motivations for the usage of 3D printers. First, pieces can be produced directly from a CAD drawing \cite{Sachs:1992cb}, which make it ideal for rapid prototyping. The time from finishing a 3D model to using it ranges from a few hours (if an in-house printer is available) to a few days (for outsourced production). Furthermore the field is rapidly evolving and recently a number of enterprises offer manufacturing at very competitive prices (usually much cheaper than milling). These advantages hold especially for small, individual pieces as gas jet nozzles. Note that the technology is not competitive in case of larger pieces and batch production.\cite{Conner:2014hr}

Last, additive manufacturing can offer more freedom in the design than conventional milling does. However, in order for this argument to hold, we need to respect some guidelines, which we are going to discuss in this paper. As readers from the laser and accelerator communities may not be familiar with additive manufacturing, we will briefly introduce the most common consumer and commercial technologies.

\subsection{Fused deposition modeling (FDM)}
Fused deposition modeling (FDM) was invented in the 1980s by Scott S. Crump \cite{Crump:1992ti}. The basic concept of FDM is similar to a common hot glue gun: A thermoplastic filament is fed through a hot nozzle. Due to the heating the filament melts and is extruded. When the material is 'printed' onto the underlying layer of material it cools down and solidifies. Due to their conceptional simplicity FDM printers are used in most consumer printers (MakerBot, Printrbot, etc.) as well as for free and open source hardware projects like RepRap or Fab@Home. Such printers are low-cost and easily affordable even for small laboratories and research groups. For our study we have used a commercial FDM printer (HP DesignJet 3D, now relabeled uPrint SE), which prints at a layer thickness of 0.254 mm.

\subsection{Stereolithography (SLA)}

Stereolithography (SLA) \cite{Hull:1986vi} is a technique used in many commercial printers, based on induced polymerization by light \cite{Jacobs:1992wg}. Here a liquid photopolymer (resin) is printed onto a surface and hardens when it is illuminated by a UV laser. In contrast to FDM machines most SLA printers have no moving head and instead the laser scans over the liquid polymer bath using a scanning mirror. Once the laser has scanned one entire layer, a leveling blade passes to smoothen the surface. Also SLA printers need to create a support structure that is removed after printing. 

For this study we have relied on PolyJet, a particular type of SLA, developed by Stratasys. Here the photopolymer is delivered in form of droplets which are immediately solidified by a UV light in the printer head. In principle the resolution of the technique is limited by the droplet size and not by the laser spot size as conventional SLA. The Stratasys Objet30 Pro printer, which we used in this study, has a layer thickness of 28 $\mu$m and an xy resolution of 42 micrometers. 
In principle, this technique should lead to the highest resolution (see discussion in following sections).

\subsection{Selective laser sintering (SLS)}

In selective laser sintering (SLS) \cite{Deckard:1989wj} a high power laser, typically a pulsed CO$_2$ or ND:YAG laser, is focussed on a powder that is then locally fused via sintering \cite{Kruth:2005hv}. The powder can be a metal compound or a polymer, e.g. nylon. Once a whole layer has been selectively sintered, the recoater applies a new layer of powder is applied on top and the procedure repeats. A main difference to FDM and SLA is that SLS does not require any support structures as the un-sintered powder itself acts as support.

SLS systems are much more expensive than FDM and SLA systems as they require high power lasers for the sintering process. In this study we used an EOS Formiga P110 printer, which uses a 30 W CO$_2$ laser coupled to an F-theta scanning lens. The printer material is a PA 12 based fine polyamide (PA 2200) with an average grain size of 60 $\mu$m. The layer thickness is 60 - 150 $\mu$m.

\section{3D printing for laser-plasma accelerators}

The use of 3D printing for laser wakefield accelerators has been pioneered at University of Michigan, where targets, especially gas cells, were produced using SLA \cite{Vargas:2014dg}. The group presented very promising results, e.g. 100 $\mu$m diameter nozzles \cite{Jolly:2012ep}. However, these parts required post-processing of the nozzles because the narrowest part of the nozzle was blocked. 

We have tested FDM, SLA and SLS printers and experienced similar issues, even at diameters exceeding $>500$ $\mu$m. When trying to address these issues with the commercial manufacturer, we were advised to avoid channels smaller than 2 mm. However, such basic guidelines are oriented to the wide public and usually differ substantially from the requirements for this special application. This lack of specific documentation has also been acknowledged in the mechanical engineering community \cite{Hernandez:2015wc}.

During our first tests we found that the quality of the printed pieces depends on the nozzle height, aperture and the print direction. In order to evaluate the actual limitations of the systems we have performed a more systematic evaluation, that involved the basic shapes that form a gas jet nozzle, i.e. tubes and hollow cones. As an example a study of different cone opening angles (a) using a transparent resin in shown in Fig.3a. 

The main design failure is when a gas jet gets blocked, so we especially investigated the influence performance of the printers for tubes of different sizes. We therefore conceived a test object consisting of holes of 0.1 - 5 mm diameter and 1, 2, 5 and 10 millimeter thickness (Fig.3d). Backlighted photographies, taken with a Canon EOS 600D digital camera, for both SLA and SLS are shown in Fig.3b/c. Using the backlighting it is evident which holes are blocked with material and therefore unsuitable for gas jet design. We observe not only a diameter dependence, but also strong influence of the channel depth, especially in the SLA case. At 1mm thickness the SLS printed piece is open up to diameters of 0.9 mm, at 2mm already holes of 1mm are not always reproducible and above only holes with diameters greater than 1.5 mm remain open. For SLA we observe an even stronger  dependence on the aspect ratio: While thin samples remain open for diameters $\geq0.7$ mm, the pieces get quickly stuck the thicker the sample. For a 10 mm channel only samples with an opening $\geq3$ mm remain unblocked. 

\begin{figure}[t]
\centering
\includegraphics[width=0.98\linewidth]{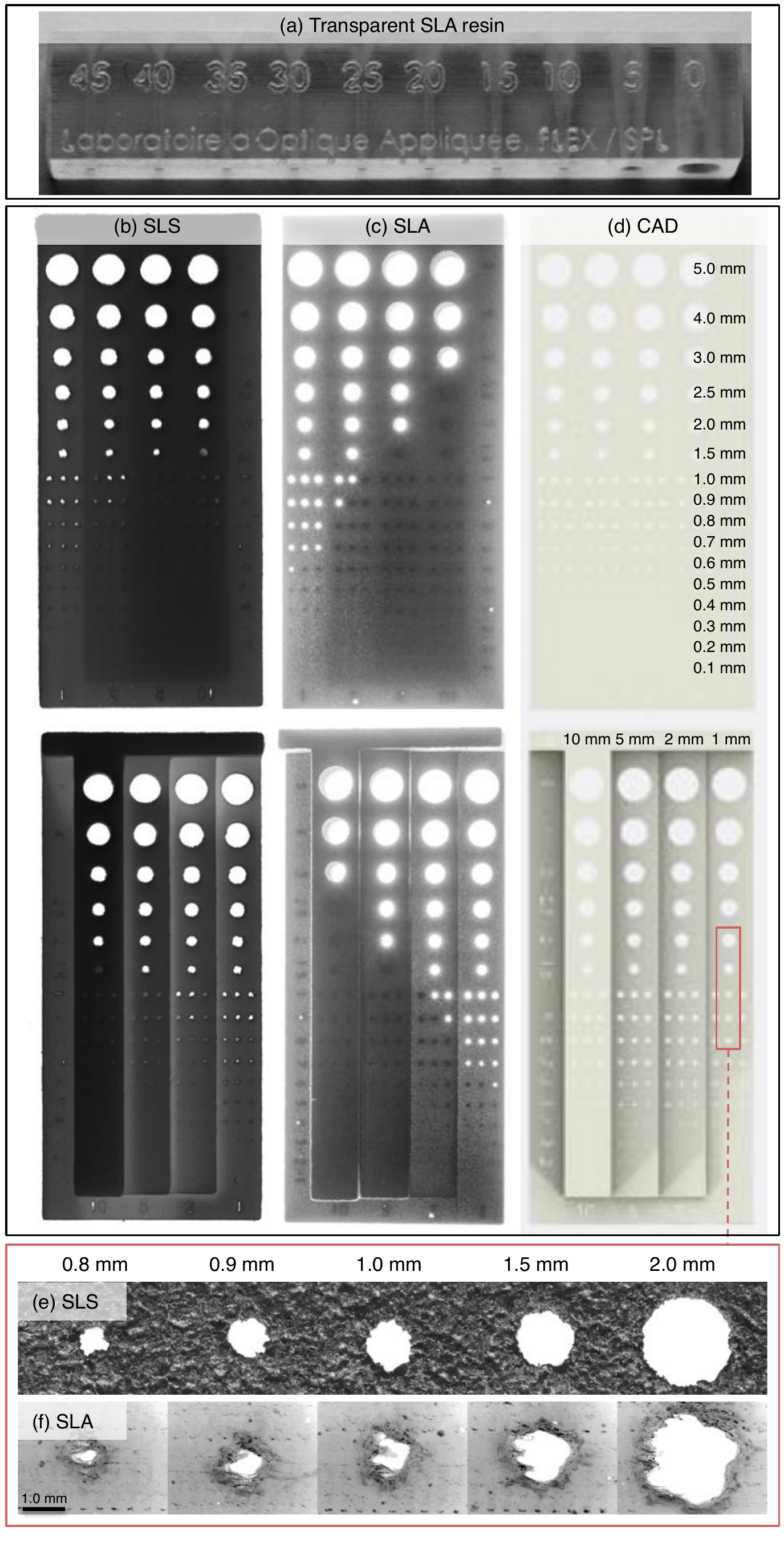}
\caption{(a) Example for a cone test print with different opening angles $\varphi$. (b-f) Test prints to determine the effective resolution of SLA and SLS printers. (d) is a render of the original CAD model, with holes from 0.1 to 5.0 mm diameter and depth from 1 to 10 mm. (b) and (c) show backlighted photographies of the front and back of the test piece. (e) and (f) show microscopy images for a selection of holes marked by the red rectangle in (d).}
\label{fig3}
\end{figure}

We also used a Zeiss Axio Imager.A2m microscope to look in detail at the quality of the tube openings for 1mm depth and 0.8 to 2.0 mm diameter (Fig.3e/f). It is obvious that SLS (e) and SLA (f) have a very different performance. The granular SLS reproduces well the circular shape at 2 mm diameter, but struggles with 0.8 mm. So while it seemed in Fig.3c that SLA performs well for thin samples, the microscopy images show that the shape is not well reproduced.

\begin{figure}[t]
\centering
\includegraphics[width=1.0\linewidth]{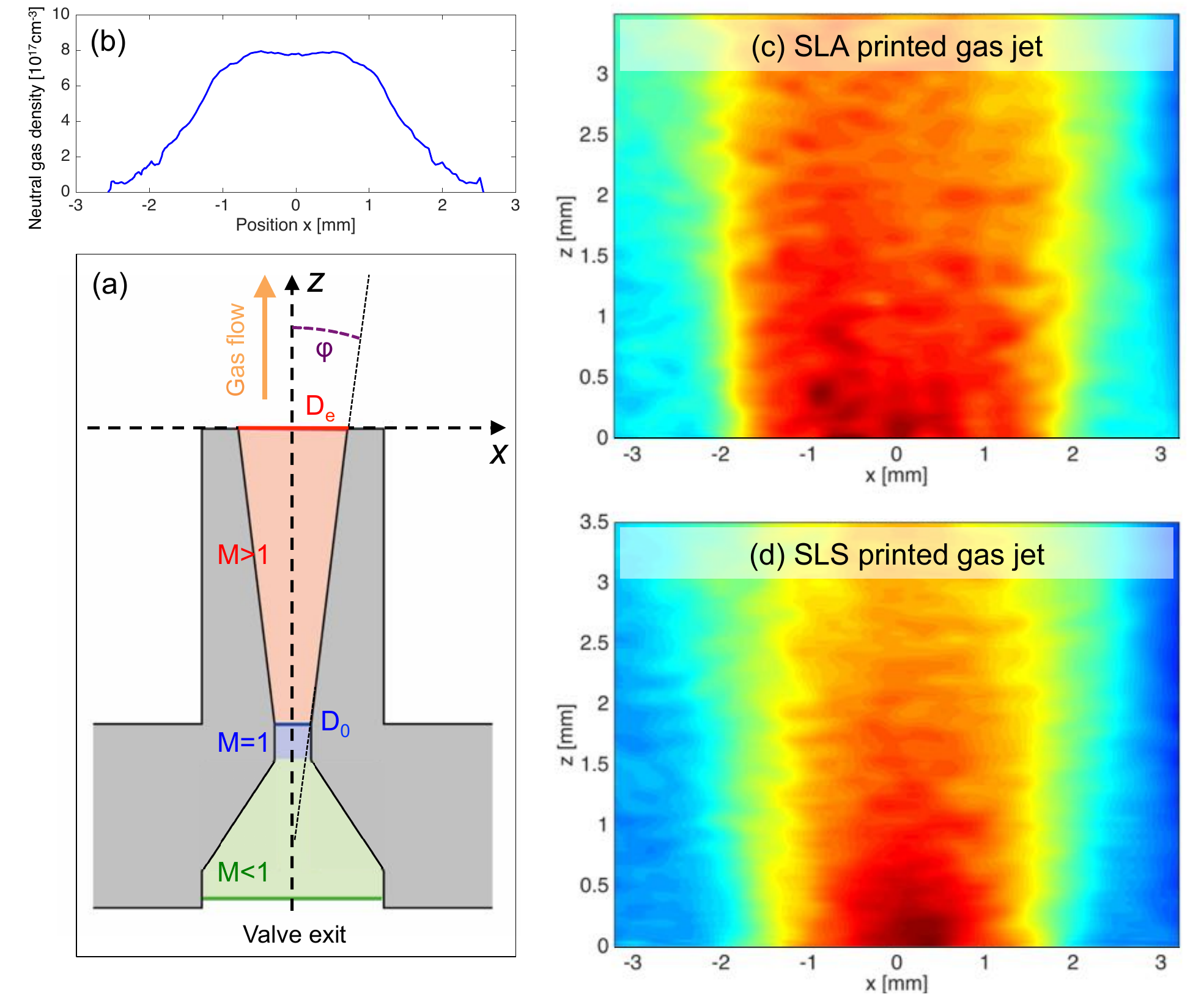}
\caption{(a) Schematic design of a supersonic de Laval nozzle. Main parameters are the inner diameter $D_0$ and the exit diameter $D_e$, the Mach number $M$ and the opening angle $\phi$. (b) Interferometrically measured density profiles at $z=0.3$ mm for a FDM printed jet. (c) and (d) show density profile maps $n_e(x,z)$ for SLA and SLS printed nozzles, respectively.}
\label{fig4}
\end{figure}

Figure 4a shows the schematics of a standard nozzle design. We created a set of test nozzles for exit diameters $D_e=1- 3$ mm, different Mach numbers $M\simeq D_e/D_0=2-4$ and opening angles $\varphi=10-30^{\circ}$. The total set consists of 45 different nozzles heads for each SLA and SLS. Again we find that only nozzles with diameters greater than 2 mm can be printed without need of post-processing. As an example Fig.4 (c-d) shows interferometric measurements of the gas flow from Mach 3 nozzles of 3 mm diameter. The measurements are taken in vacuum with a HeNe laser illuminating the sample. The beam is then self-interfered using a Nomarski interferometer \cite{Benattar:1979uha}. Density values are estimated from the phase shift assuming circular symmetry that allows us to apply an Abel transform. While the exit hole $D_e$ is better reproduced by SLS than SLA (see Fig.3e/f), we observe that the SLA gas flow is more homogeneous than the SLS jet. This is due to the fact that SLS, though leading to a less regular exit hole shape, has a better performance of printing small tubes of several millimeter length as seen in Fig.3b/c. This results in an overall better printing of the entrance hole $D_0$ than for the SLS case. Also, the resin has a reduced surface roughness compared to the grains used for SLA, which should also increase the Mach number in favor of SLS.

\begin{figure}[t]
\centering
\includegraphics[width=1.0\linewidth]{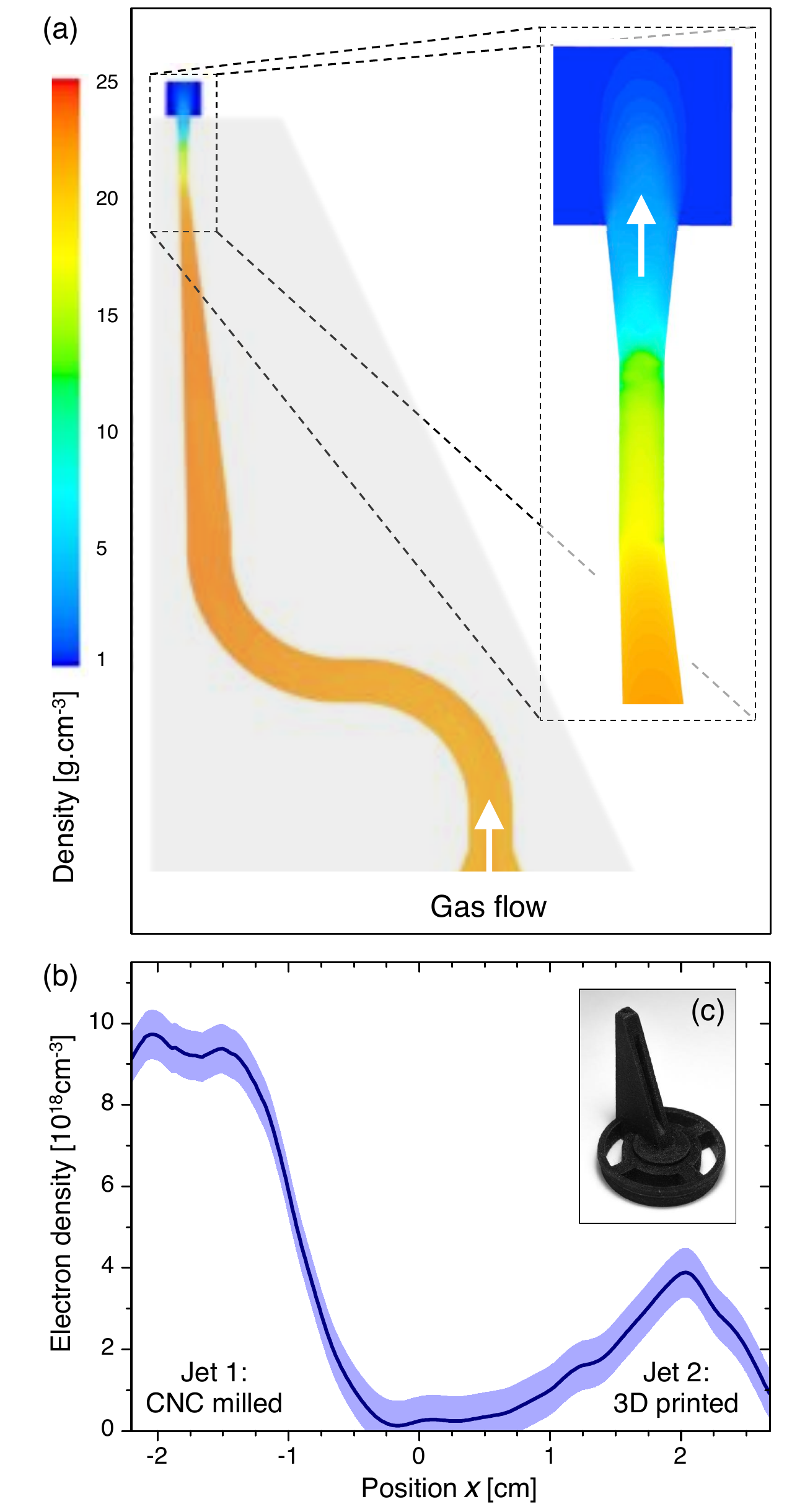}
\caption{(a) \textsc{Ansys Fluent} simulation of the density profile in an asymmetric gas jet nozzle. As seen in the enlarged frame the exit flow is mostly symmetric, despite the asymmetrical layout. (b) Interferometric gas density measurements along the laser propagation from experiments on laser-plasma lensing \cite{Thaury:2015cg}. Due to the sonic expansion the gas profile of the 3D printed jet is peaked, while the supersonic flow of the adjacent aluminum nozzle leads to a sharper density transition. The final printed nozzle design is shown as inset (c).}
\label{fig5}
\end{figure}

\section{Application in experiments}

After the initial characterization from the preceding section, we have employed 3D printed gas jet nozzles in laser wakefield acceleration experiments. For this we used the \textsc{Salle Jaune} Laser at Laboratoire d'Optique Appliqu\'ee. The system delivers linearly polarized 1 J / 28 fs (FWHM) pulses at a central wavelength of 810 nm, which are then focussed onto the gas target with an off-axis parabola. As reservoir gas we have used pure helium.

In a first experiment we have used the FDM printed jets (Fig.4b) to accelerate self-injected electrons \cite{Corde:2013gj} to energies in the 100-300 MeV range. As the results of this experiment will be published elsewhere, we will focus our discussion on the performance of the jet. First of all, we have not experienced significant performance differences between 3D printed and CNC milled nozzles. During test the nozzles easily withstood backing pressures of up to 50 bar. Also, while the laser-plasma interaction lead to visible degradation of the surface,  performance of the jet showed no significant change over several hundred laser shots. Furthermore it was reported that 3D printed materials suffer from significant outgasing \cite{Povilus:2014jz}. However, for the experimental requirements of laser wakefield acceleration (vacuum pressures of $\sim 10^{-3}-10^{-5}$ mbar), this is no problem and 3D printed parts can be used without concerns.

One of the main premises of 3D printing is the possibility of creating complex geometries. Such geometries can be required for advanced experimental configuration, like a laser-plasma accelerator coupled to a laser-plasma lens \cite{Thaury:2015cg}. Here a single laser pulse serves as driver for the wakefield accelerator, but also creates focusing fields in a subsequent jet. This second jet needs to be placed just a few millimeters behind the exit of the first. The jets cannot be placed opposing to each other, as this would cause turbulences in the gas flow. Instead, the nozzle exit has to be placed close to the outer edge of the valve. Using traditional production techniques such jets would be produced using either milling of two separate pieces or by molding. The former requires high accuracy in production and alignment, while molding requires a core to be produced and micro-molding is rather costly. As an alternative we have designed a new gas jet nozzle whose exit is displace with respect to the vale. The initial CAD model was imported into the computational fluid dynamics software \textsc{Ansys Fluent}, where we simulated the gas expansion into vacuum on a 2D adaptive mesh. As shown in Fig.5a there is little influence of the design on the flow symmetry at the exit. We therefore went on to print the nozzle using SLS and SLA. As expected from our previous study the deep channel was blocked for SLS printing. Instead we chose the SLA nozzle and post processed the exit with a drill.  The nozzle was then used in the experiment on relativistic electron beam focusing using a laser-plasma lens. While the design could not provide a sharp density profile due to the sonic flow, it allowed to place the nozzle very close to the exit of a first gas jet and thus serve as focusing element for the electron beam \cite{Thaury:2015cg}.

\section{Conclusions and Outlook}

In conclusion we have presented the current state of our ongoing investigations on various 3D printing (additive manufacturing) techniques for the production of gas jet nozzles. Using commercial printers we estimate as a current design guideline to print at least with a minimal diameter of 1mm in order to avoid obstructions in the gas channel. As seen in Fig.1 this will be sufficient for most experiments on laser wakefield acceleration. We also demonstrated that 3D printing allows the design of unconventional nozzle designs for special applications, e.g. as second nozzle for a laser-plasma lens.
As 3D printed parts possess a good structural integrity (except sheering forces along the print layer), we have also started to investigate further usage in experiment. For instance, the technique is well-suited to design and timely produce personalized mounts. Also, for certain materials an additional surface treatment, e.g. with acetone, may serve as simple solution to improve the print quality.
Including the continuing prize reduction, availability of new materials and so forth, we anticipate that the technology will soon find its way into more laboratory applications. \\

ACKNOWLEDGMENTS: The authors acknowledge support from the Agence Nationale pour la Recherche through the FENICS Project No. ANR-12-JS04-0004-01, the Agence Nationale pour la Recherche through the FEMTOMAT Project No. ANR-13-BS04-0002, the X-Five ERC project (Contract No. 339128), EuCARD2/ANAC2 EC FP7 project (Contract No. 312453), LA3NET project (GA-ITN-2011-289191), and GARC project 15-03118S. A.D. acknowledges S. Zierke (RWTH Aachen University), F. Sylla (SourceLab) and C. Ruiz (Universidad de Salamanca) for helpful discussions, and FabLab (Institut d'Optique) and Sculpteo for printing services and support.

\bibliographystyle{apsrev4-1}

\begin{thebibliography}{45}%
\makeatletter
\providecommand \@ifxundefined [1]{%
 \@ifx{#1\undefined}
}%
\providecommand \@ifnum [1]{%
 \ifnum #1\expandafter \@firstoftwo
 \else \expandafter \@secondoftwo
 \fi
}%
\providecommand \@ifx [1]{%
 \ifx #1\expandafter \@firstoftwo
 \else \expandafter \@secondoftwo
 \fi
}%
\providecommand \natexlab [1]{#1}%
\providecommand \enquote  [1]{``#1''}%
\providecommand \bibnamefont  [1]{#1}%
\providecommand \bibfnamefont [1]{#1}%
\providecommand \citenamefont [1]{#1}%
\providecommand \href@noop [0]{\@secondoftwo}%
\providecommand \href [0]{\begingroup \@sanitize@url \@href}%
\providecommand \@href[1]{\@@startlink{#1}\@@href}%
\providecommand \@@href[1]{\endgroup#1\@@endlink}%
\providecommand \@sanitize@url [0]{\catcode `\\12\catcode `\$12\catcode
  `\&12\catcode `\#12\catcode `\^12\catcode `\_12\catcode `\%12\relax}%
\providecommand \@@startlink[1]{}%
\providecommand \@@endlink[0]{}%
\providecommand \url  [0]{\begingroup\@sanitize@url \@url }%
\providecommand \@url [1]{\endgroup\@href {#1}{\urlprefix }}%
\providecommand \urlprefix  [0]{URL }%
\providecommand \Eprint [0]{\href }%
\providecommand \doibase [0]{http://dx.doi.org/}%
\providecommand \selectlanguage [0]{\@gobble}%
\providecommand \bibinfo  [0]{\@secondoftwo}%
\providecommand \bibfield  [0]{\@secondoftwo}%
\providecommand \translation [1]{[#1]}%
\providecommand \BibitemOpen [0]{}%
\providecommand \bibitemStop [0]{}%
\providecommand \bibitemNoStop [0]{.\EOS\space}%
\providecommand \EOS [0]{\spacefactor3000\relax}%
\providecommand \BibitemShut  [1]{\csname bibitem#1\endcsname}%
\let\auto@bib@innerbib\@empty
\bibitem [{\citenamefont {Wiedemann}(2015)}]{Wiedemann:2015ws}%
  \BibitemOpen
  \bibfield  {author} {\bibinfo {author} {\bibfnamefont {H.}~\bibnamefont
  {Wiedemann}},\ }\href@noop {} {\emph {\bibinfo {title} {{Particle Accelerator
  Physics}}}},\ \bibinfo {edition} {4th}\ ed.\ (\bibinfo  {publisher} {Springer
  International Publishing},\ \bibinfo {year} {2015})\BibitemShut {NoStop}%
\bibitem [{\citenamefont {Esarey}\ \emph {et~al.}(2009)\citenamefont {Esarey},
  \citenamefont {Schroeder},\ and\ \citenamefont {Leemans}}]{Esarey:2009ks}%
  \BibitemOpen
  \bibfield  {author} {\bibinfo {author} {\bibfnamefont {E.}~\bibnamefont
  {Esarey}}, \bibinfo {author} {\bibfnamefont {C.~B.}\ \bibnamefont
  {Schroeder}}, \ and\ \bibinfo {author} {\bibfnamefont {W.~P.}\ \bibnamefont
  {Leemans}},\ }\href@noop {} {\bibfield  {journal} {\bibinfo  {journal}
  {Reviews of Modern Physics}\ }\textbf {\bibinfo {volume} {81}},\ \bibinfo
  {pages} {1229} (\bibinfo {year} {2009})}\BibitemShut {NoStop}%
\bibitem [{\citenamefont {Malka}\ \emph {et~al.}(2002)\citenamefont {Malka},
  \citenamefont {Fritzler}, \citenamefont {Lefebvre}, \citenamefont {Aleonard},
  \citenamefont {Burgy}, \citenamefont {Chambaret}, \citenamefont {Chemin},
  \citenamefont {Krushelnick}, \citenamefont {Malka},\ and\ \citenamefont
  {Mangles}}]{Malka:2002eu}%
  \BibitemOpen
  \bibfield  {author} {\bibinfo {author} {\bibfnamefont {V.}~\bibnamefont
  {Malka}}, \bibinfo {author} {\bibfnamefont {S.}~\bibnamefont {Fritzler}},
  \bibinfo {author} {\bibfnamefont {E.}~\bibnamefont {Lefebvre}}, \bibinfo
  {author} {\bibfnamefont {M.-M.}\ \bibnamefont {Aleonard}}, \bibinfo {author}
  {\bibfnamefont {F.}~\bibnamefont {Burgy}}, \bibinfo {author} {\bibfnamefont
  {J.~P.}\ \bibnamefont {Chambaret}}, \bibinfo {author} {\bibfnamefont {J.-F.}\
  \bibnamefont {Chemin}}, \bibinfo {author} {\bibfnamefont {K.}~\bibnamefont
  {Krushelnick}}, \bibinfo {author} {\bibfnamefont {G.}~\bibnamefont {Malka}},
  \ and\ \bibinfo {author} {\bibfnamefont {S.}~\bibnamefont {Mangles}},\
  }\href@noop {} {\bibfield  {journal} {\bibinfo  {journal} {Science (New York,
  NY)}\ }\textbf {\bibinfo {volume} {298}},\ \bibinfo {pages} {1596} (\bibinfo
  {year} {2002})}\BibitemShut {NoStop}%
\bibitem [{\citenamefont {Glinec}\ \emph {et~al.}(2005)\citenamefont {Glinec},
  \citenamefont {Faure}, \citenamefont {Dain}, \citenamefont {Darbon},
  \citenamefont {Hosokai}, \citenamefont {Santos}, \citenamefont {Lefebvre},
  \citenamefont {Rousseau}, \citenamefont {Burgy}, \citenamefont {Mercier},\
  and\ \citenamefont {Malka}}]{Glinec:2005ve}%
  \BibitemOpen
  \bibfield  {author} {\bibinfo {author} {\bibfnamefont {Y.}~\bibnamefont
  {Glinec}}, \bibinfo {author} {\bibfnamefont {J.}~\bibnamefont {Faure}},
  \bibinfo {author} {\bibfnamefont {L.~L.}\ \bibnamefont {Dain}}, \bibinfo
  {author} {\bibfnamefont {S.}~\bibnamefont {Darbon}}, \bibinfo {author}
  {\bibfnamefont {T.}~\bibnamefont {Hosokai}}, \bibinfo {author} {\bibfnamefont
  {J.~J.}\ \bibnamefont {Santos}}, \bibinfo {author} {\bibfnamefont
  {E.}~\bibnamefont {Lefebvre}}, \bibinfo {author} {\bibfnamefont {J.~P.}\
  \bibnamefont {Rousseau}}, \bibinfo {author} {\bibfnamefont {F.}~\bibnamefont
  {Burgy}}, \bibinfo {author} {\bibfnamefont {B.}~\bibnamefont {Mercier}}, \
  and\ \bibinfo {author} {\bibfnamefont {V.}~\bibnamefont {Malka}},\
  }\href@noop {} {\bibfield  {journal} {\bibinfo  {journal} {Physical Review
  Letters}\ }\textbf {\bibinfo {volume} {94}},\ \bibinfo {pages} {025003}
  (\bibinfo {year} {2005})}\BibitemShut {NoStop}%
\bibitem [{\citenamefont {D{\"o}pp}\ \emph
  {et~al.}(2016{\natexlab{a}})\citenamefont {D{\"o}pp}, \citenamefont
  {Guillaume}, \citenamefont {Thaury}, \citenamefont {Lifschitz}, \citenamefont
  {Sylla}, \citenamefont {Goddet}, \citenamefont {Tafzi}, \citenamefont
  {Iaquanello}, \citenamefont {Lefrou}, \citenamefont {Rousseau}, \citenamefont
  {Conejero}, \citenamefont {Ruiz}, \citenamefont {Ta~Phuoc},\ and\
  \citenamefont {Malka}}]{Dopp:2016tv}%
  \BibitemOpen
  \bibfield  {author} {\bibinfo {author} {\bibfnamefont {A.}~\bibnamefont
  {D{\"o}pp}}, \bibinfo {author} {\bibfnamefont {E.}~\bibnamefont {Guillaume}},
  \bibinfo {author} {\bibfnamefont {C.}~\bibnamefont {Thaury}}, \bibinfo
  {author} {\bibfnamefont {A.}~\bibnamefont {Lifschitz}}, \bibinfo {author}
  {\bibfnamefont {F.}~\bibnamefont {Sylla}}, \bibinfo {author} {\bibfnamefont
  {J.~P.}\ \bibnamefont {Goddet}}, \bibinfo {author} {\bibfnamefont
  {A.}~\bibnamefont {Tafzi}}, \bibinfo {author} {\bibfnamefont
  {G.}~\bibnamefont {Iaquanello}}, \bibinfo {author} {\bibfnamefont
  {T.}~\bibnamefont {Lefrou}}, \bibinfo {author} {\bibfnamefont
  {P.}~\bibnamefont {Rousseau}}, \bibinfo {author} {\bibfnamefont
  {E.}~\bibnamefont {Conejero}}, \bibinfo {author} {\bibfnamefont
  {C.}~\bibnamefont {Ruiz}}, \bibinfo {author} {\bibfnamefont {K.}~\bibnamefont
  {Ta~Phuoc}}, \ and\ \bibinfo {author} {\bibfnamefont {V.}~\bibnamefont
  {Malka}},\ }\href@noop {} {\bibfield  {journal} {\bibinfo  {journal} {Nuclear
  Instruments {\&} Methods in Physics Research A}\ } (\bibinfo {year}
  {2016}{\natexlab{a}})}\BibitemShut {NoStop}%
\bibitem [{\citenamefont {Schlenvoigt}\ \emph {et~al.}(2008)\citenamefont
  {Schlenvoigt}, \citenamefont {Haupt}, \citenamefont {Debus}, \citenamefont
  {Budde}, \citenamefont {Jackel}, \citenamefont {Pfotenhauer}, \citenamefont
  {Schwoerer}, \citenamefont {Rohwer}, \citenamefont {Gallacher}, \citenamefont
  {Brunetti}, \citenamefont {Shanks}, \citenamefont {Wiggins},\ and\
  \citenamefont {Jaroszynski}}]{Schlenvoigt:2008bg}%
  \BibitemOpen
  \bibfield  {author} {\bibinfo {author} {\bibfnamefont {H.~P.}\ \bibnamefont
  {Schlenvoigt}}, \bibinfo {author} {\bibfnamefont {K.}~\bibnamefont {Haupt}},
  \bibinfo {author} {\bibfnamefont {A.}~\bibnamefont {Debus}}, \bibinfo
  {author} {\bibfnamefont {F.}~\bibnamefont {Budde}}, \bibinfo {author}
  {\bibfnamefont {O.}~\bibnamefont {Jackel}}, \bibinfo {author} {\bibfnamefont
  {S.}~\bibnamefont {Pfotenhauer}}, \bibinfo {author} {\bibfnamefont
  {H.}~\bibnamefont {Schwoerer}}, \bibinfo {author} {\bibfnamefont
  {E.}~\bibnamefont {Rohwer}}, \bibinfo {author} {\bibfnamefont {J.~G.}\
  \bibnamefont {Gallacher}}, \bibinfo {author} {\bibfnamefont {E.}~\bibnamefont
  {Brunetti}}, \bibinfo {author} {\bibfnamefont {R.~P.}\ \bibnamefont
  {Shanks}}, \bibinfo {author} {\bibfnamefont {S.~M.}\ \bibnamefont {Wiggins}},
  \ and\ \bibinfo {author} {\bibfnamefont {D.~A.}\ \bibnamefont
  {Jaroszynski}},\ }\href@noop {} {\bibfield  {journal} {\bibinfo  {journal}
  {Nature Physics}\ }\textbf {\bibinfo {volume} {4}},\ \bibinfo {pages} {130}
  (\bibinfo {year} {2008})}\BibitemShut {NoStop}%
\bibitem [{\citenamefont {Ta~Phuoc}\ \emph {et~al.}(2012)\citenamefont
  {Ta~Phuoc}, \citenamefont {Corde}, \citenamefont {Thaury}, \citenamefont
  {Malka}, \citenamefont {Tafzi}, \citenamefont {Goddet}, \citenamefont {Shah},
  \citenamefont {Sebban},\ and\ \citenamefont {Rousse}}]{TaPhuoc:2012cg}%
  \BibitemOpen
  \bibfield  {author} {\bibinfo {author} {\bibfnamefont {K.}~\bibnamefont
  {Ta~Phuoc}}, \bibinfo {author} {\bibfnamefont {S.}~\bibnamefont {Corde}},
  \bibinfo {author} {\bibfnamefont {C.}~\bibnamefont {Thaury}}, \bibinfo
  {author} {\bibfnamefont {V.}~\bibnamefont {Malka}}, \bibinfo {author}
  {\bibfnamefont {A.}~\bibnamefont {Tafzi}}, \bibinfo {author} {\bibfnamefont
  {J.~P.}\ \bibnamefont {Goddet}}, \bibinfo {author} {\bibfnamefont {R.~C.}\
  \bibnamefont {Shah}}, \bibinfo {author} {\bibfnamefont {S.}~\bibnamefont
  {Sebban}}, \ and\ \bibinfo {author} {\bibfnamefont {A.}~\bibnamefont
  {Rousse}},\ }\href@noop {} {\bibfield  {journal} {\bibinfo  {journal} {Nature
  Photonics}\ }\textbf {\bibinfo {volume} {6}},\ \bibinfo {pages} {308}
  (\bibinfo {year} {2012})}\BibitemShut {NoStop}%
\bibitem [{\citenamefont {Powers}\ \emph {et~al.}(2013)\citenamefont {Powers},
  \citenamefont {Ghebregziabher}, \citenamefont {Golovin}, \citenamefont {Liu},
  \citenamefont {Chen}, \citenamefont {Banerjee}, \citenamefont {Zhang},\ and\
  \citenamefont {Umstadter}}]{Powers:2013bx}%
  \BibitemOpen
  \bibfield  {author} {\bibinfo {author} {\bibfnamefont {N.~D.}\ \bibnamefont
  {Powers}}, \bibinfo {author} {\bibfnamefont {I.}~\bibnamefont
  {Ghebregziabher}}, \bibinfo {author} {\bibfnamefont {G.}~\bibnamefont
  {Golovin}}, \bibinfo {author} {\bibfnamefont {C.}~\bibnamefont {Liu}},
  \bibinfo {author} {\bibfnamefont {S.}~\bibnamefont {Chen}}, \bibinfo {author}
  {\bibfnamefont {S.}~\bibnamefont {Banerjee}}, \bibinfo {author}
  {\bibfnamefont {J.}~\bibnamefont {Zhang}}, \ and\ \bibinfo {author}
  {\bibfnamefont {D.~P.}\ \bibnamefont {Umstadter}},\ }\href@noop {} {\bibfield
   {journal} {\bibinfo  {journal} {Nature Photonics}\ ,\ \bibinfo {pages} {1}}
  (\bibinfo {year} {2013})}\BibitemShut {NoStop}%
\bibitem [{\citenamefont {Rousse}\ \emph {et~al.}(2004)\citenamefont {Rousse},
  \citenamefont {Phuoc}, \citenamefont {Shah}, \citenamefont {Pukhov},
  \citenamefont {Lefebvre}, \citenamefont {Malka}, \citenamefont {Kiselev},
  \citenamefont {Burgy}, \citenamefont {Rousseau}, \citenamefont {Umstadter},\
  and\ \citenamefont {Hulin}}]{Rousse:2004tc}%
  \BibitemOpen
  \bibfield  {author} {\bibinfo {author} {\bibfnamefont {A.}~\bibnamefont
  {Rousse}}, \bibinfo {author} {\bibfnamefont {K.~T.}\ \bibnamefont {Phuoc}},
  \bibinfo {author} {\bibfnamefont {R.}~\bibnamefont {Shah}}, \bibinfo {author}
  {\bibfnamefont {A.}~\bibnamefont {Pukhov}}, \bibinfo {author} {\bibfnamefont
  {E.}~\bibnamefont {Lefebvre}}, \bibinfo {author} {\bibfnamefont
  {V.}~\bibnamefont {Malka}}, \bibinfo {author} {\bibfnamefont
  {S.}~\bibnamefont {Kiselev}}, \bibinfo {author} {\bibfnamefont
  {F.}~\bibnamefont {Burgy}}, \bibinfo {author} {\bibfnamefont {J.~P.}\
  \bibnamefont {Rousseau}}, \bibinfo {author} {\bibfnamefont {D.}~\bibnamefont
  {Umstadter}}, \ and\ \bibinfo {author} {\bibfnamefont {D.}~\bibnamefont
  {Hulin}},\ }\href@noop {} {\bibfield  {journal} {\bibinfo  {journal}
  {Physical Review Letters}\ }\textbf {\bibinfo {volume} {93}},\ \bibinfo
  {pages} {135005} (\bibinfo {year} {2004})}\BibitemShut {NoStop}%
\bibitem [{\citenamefont {Schmid}\ \emph {et~al.}(2010)\citenamefont {Schmid},
  \citenamefont {Buck}, \citenamefont {Sears}, \citenamefont {Mikhailova},
  \citenamefont {Tautz}, \citenamefont {Herrmann}, \citenamefont {Geissler},
  \citenamefont {Krausz},\ and\ \citenamefont {Veisz}}]{Schmid:2010ih}%
  \BibitemOpen
  \bibfield  {author} {\bibinfo {author} {\bibfnamefont {K.}~\bibnamefont
  {Schmid}}, \bibinfo {author} {\bibfnamefont {A.}~\bibnamefont {Buck}},
  \bibinfo {author} {\bibfnamefont {C.~M.~S.}\ \bibnamefont {Sears}}, \bibinfo
  {author} {\bibfnamefont {J.~M.}\ \bibnamefont {Mikhailova}}, \bibinfo
  {author} {\bibfnamefont {R.}~\bibnamefont {Tautz}}, \bibinfo {author}
  {\bibfnamefont {D.}~\bibnamefont {Herrmann}}, \bibinfo {author}
  {\bibfnamefont {M.}~\bibnamefont {Geissler}}, \bibinfo {author}
  {\bibfnamefont {F.}~\bibnamefont {Krausz}}, \ and\ \bibinfo {author}
  {\bibfnamefont {L.}~\bibnamefont {Veisz}},\ }\href@noop {} {\bibfield
  {journal} {\bibinfo  {journal} {Physical Review Special Topics-Accelerators
  and Beams}\ }\textbf {\bibinfo {volume} {13}},\ \bibinfo {pages} {091301}
  (\bibinfo {year} {2010})}\BibitemShut {NoStop}%
\bibitem [{\citenamefont {Buck}\ \emph {et~al.}(2013)\citenamefont {Buck},
  \citenamefont {Wenz}, \citenamefont {Xu}, \citenamefont {Khrennikov},
  \citenamefont {Schmid}, \citenamefont {Heigoldt}, \citenamefont {Mikhailova},
  \citenamefont {Geissler}, \citenamefont {Shen}, \citenamefont {Krausz},
  \citenamefont {Karsch},\ and\ \citenamefont {Veisz}}]{Buck:2013gs}%
  \BibitemOpen
  \bibfield  {author} {\bibinfo {author} {\bibfnamefont {A.}~\bibnamefont
  {Buck}}, \bibinfo {author} {\bibfnamefont {J.}~\bibnamefont {Wenz}}, \bibinfo
  {author} {\bibfnamefont {J.}~\bibnamefont {Xu}}, \bibinfo {author}
  {\bibfnamefont {K.}~\bibnamefont {Khrennikov}}, \bibinfo {author}
  {\bibfnamefont {K.}~\bibnamefont {Schmid}}, \bibinfo {author} {\bibfnamefont
  {M.}~\bibnamefont {Heigoldt}}, \bibinfo {author} {\bibfnamefont {J.~M.}\
  \bibnamefont {Mikhailova}}, \bibinfo {author} {\bibfnamefont
  {M.}~\bibnamefont {Geissler}}, \bibinfo {author} {\bibfnamefont
  {B.}~\bibnamefont {Shen}}, \bibinfo {author} {\bibfnamefont {F.}~\bibnamefont
  {Krausz}}, \bibinfo {author} {\bibfnamefont {S.}~\bibnamefont {Karsch}}, \
  and\ \bibinfo {author} {\bibfnamefont {L.}~\bibnamefont {Veisz}},\
  }\href@noop {} {\bibfield  {journal} {\bibinfo  {journal} {Physical Review
  Letters}\ }\textbf {\bibinfo {volume} {110}},\ \bibinfo {pages} {185006}
  (\bibinfo {year} {2013})}\BibitemShut {NoStop}%
\bibitem [{\citenamefont {D{\"o}pp}\ \emph
  {et~al.}(2016{\natexlab{b}})\citenamefont {D{\"o}pp}, \citenamefont
  {Guillaume}, \citenamefont {Thaury}, \citenamefont {Lifschitz}, \citenamefont
  {Ta~Phuoc},\ and\ \citenamefont {Malka}}]{Dopp:2016gm}%
  \BibitemOpen
  \bibfield  {author} {\bibinfo {author} {\bibfnamefont {A.}~\bibnamefont
  {D{\"o}pp}}, \bibinfo {author} {\bibfnamefont {E.}~\bibnamefont {Guillaume}},
  \bibinfo {author} {\bibfnamefont {C.}~\bibnamefont {Thaury}}, \bibinfo
  {author} {\bibfnamefont {A.}~\bibnamefont {Lifschitz}}, \bibinfo {author}
  {\bibfnamefont {K.}~\bibnamefont {Ta~Phuoc}}, \ and\ \bibinfo {author}
  {\bibfnamefont {V.}~\bibnamefont {Malka}},\ }\href@noop {} {\bibfield
  {journal} {\bibinfo  {journal} {Physics of Plasmas}\ }\textbf {\bibinfo
  {volume} {23}},\ \bibinfo {pages} {056702} (\bibinfo {year}
  {2016}{\natexlab{b}})}\BibitemShut {NoStop}%
\bibitem [{\citenamefont {Guillaume}\ \emph {et~al.}(2015)\citenamefont
  {Guillaume}, \citenamefont {D{\"o}pp}, \citenamefont {Thaury}, \citenamefont
  {Ta~Phuoc}, \citenamefont {Lifschitz}, \citenamefont {Grittani},
  \citenamefont {Goddet}, \citenamefont {Tafzi}, \citenamefont {Chou},
  \citenamefont {Veisz},\ and\ \citenamefont {Malka}}]{Guillaume:2015dia}%
  \BibitemOpen
  \bibfield  {author} {\bibinfo {author} {\bibfnamefont {E.}~\bibnamefont
  {Guillaume}}, \bibinfo {author} {\bibfnamefont {A.}~\bibnamefont {D{\"o}pp}},
  \bibinfo {author} {\bibfnamefont {C.}~\bibnamefont {Thaury}}, \bibinfo
  {author} {\bibfnamefont {K.}~\bibnamefont {Ta~Phuoc}}, \bibinfo {author}
  {\bibfnamefont {A.}~\bibnamefont {Lifschitz}}, \bibinfo {author}
  {\bibfnamefont {G.}~\bibnamefont {Grittani}}, \bibinfo {author}
  {\bibfnamefont {J.~P.}\ \bibnamefont {Goddet}}, \bibinfo {author}
  {\bibfnamefont {A.}~\bibnamefont {Tafzi}}, \bibinfo {author} {\bibfnamefont
  {S.~W.}\ \bibnamefont {Chou}}, \bibinfo {author} {\bibfnamefont
  {L.}~\bibnamefont {Veisz}}, \ and\ \bibinfo {author} {\bibfnamefont
  {V.}~\bibnamefont {Malka}},\ }\href@noop {} {\bibfield  {journal} {\bibinfo
  {journal} {Physical Review Letters}\ }\textbf {\bibinfo {volume} {115}},\
  \bibinfo {pages} {155002} (\bibinfo {year} {2015})}\BibitemShut {NoStop}%
\bibitem [{\citenamefont {Thaury}\ \emph {et~al.}(2015)\citenamefont {Thaury},
  \citenamefont {Guillaume}, \citenamefont {D{\"o}pp}, \citenamefont {Lehe},
  \citenamefont {Lifschitz}, \citenamefont {Phuoc}, \citenamefont {Gautier},
  \citenamefont {Goddet}, \citenamefont {Tafzi}, \citenamefont {Flacco},
  \citenamefont {Tissandier}, \citenamefont {Sebban}, \citenamefont {Rousse},\
  and\ \citenamefont {Malka}}]{Thaury:2015cg}%
  \BibitemOpen
  \bibfield  {author} {\bibinfo {author} {\bibfnamefont {C.}~\bibnamefont
  {Thaury}}, \bibinfo {author} {\bibfnamefont {E.}~\bibnamefont {Guillaume}},
  \bibinfo {author} {\bibfnamefont {A.}~\bibnamefont {D{\"o}pp}}, \bibinfo
  {author} {\bibfnamefont {R.}~\bibnamefont {Lehe}}, \bibinfo {author}
  {\bibfnamefont {A.}~\bibnamefont {Lifschitz}}, \bibinfo {author}
  {\bibfnamefont {K.~T.}\ \bibnamefont {Phuoc}}, \bibinfo {author}
  {\bibfnamefont {J.}~\bibnamefont {Gautier}}, \bibinfo {author} {\bibfnamefont
  {J.~P.}\ \bibnamefont {Goddet}}, \bibinfo {author} {\bibfnamefont
  {A.}~\bibnamefont {Tafzi}}, \bibinfo {author} {\bibfnamefont
  {A.}~\bibnamefont {Flacco}}, \bibinfo {author} {\bibfnamefont
  {F.}~\bibnamefont {Tissandier}}, \bibinfo {author} {\bibfnamefont
  {S.}~\bibnamefont {Sebban}}, \bibinfo {author} {\bibfnamefont
  {A.}~\bibnamefont {Rousse}}, \ and\ \bibinfo {author} {\bibfnamefont
  {V.}~\bibnamefont {Malka}},\ }\href@noop {} {\bibfield  {journal} {\bibinfo
  {journal} {Nature Communications}\ }\textbf {\bibinfo {volume} {6}},\
  \bibinfo {pages} {1} (\bibinfo {year} {2015})}\BibitemShut {NoStop}%
\bibitem [{\citenamefont {Mangles}(2016)}]{Mangles:2016wp}%
  \BibitemOpen
  \bibfield  {author} {\bibinfo {author} {\bibfnamefont {S.}~\bibnamefont
  {Mangles}},\ }\href@noop {} {\bibfield  {journal} {\bibinfo  {journal} {CERN
  Yellow Reports}\ }\textbf {\bibinfo {volume} {1}},\ \bibinfo {pages} {289}
  (\bibinfo {year} {2016})}\BibitemShut {NoStop}%
\bibitem [{\citenamefont {Benedetti}\ \emph {et~al.}(2013)\citenamefont
  {Benedetti}, \citenamefont {Schroeder}, \citenamefont {Esarey}, \citenamefont
  {Rossi},\ and\ \citenamefont {Leemans}}]{Benedetti:2013fy}%
  \BibitemOpen
  \bibfield  {author} {\bibinfo {author} {\bibfnamefont {C.}~\bibnamefont
  {Benedetti}}, \bibinfo {author} {\bibfnamefont {C.~B.}\ \bibnamefont
  {Schroeder}}, \bibinfo {author} {\bibfnamefont {E.}~\bibnamefont {Esarey}},
  \bibinfo {author} {\bibfnamefont {F.}~\bibnamefont {Rossi}}, \ and\ \bibinfo
  {author} {\bibfnamefont {W.~P.}\ \bibnamefont {Leemans}},\ }\href@noop {}
  {\bibfield  {journal} {\bibinfo  {journal} {Physics of Plasmas}\ }\textbf
  {\bibinfo {volume} {20}},\ \bibinfo {pages} {103108} (\bibinfo {year}
  {2013})}\BibitemShut {NoStop}%
\bibitem [{\citenamefont {Butler}\ \emph {et~al.}(2002)\citenamefont {Butler},
  \citenamefont {Spence},\ and\ \citenamefont {Hooker}}]{Butler:2002ho}%
  \BibitemOpen
  \bibfield  {author} {\bibinfo {author} {\bibfnamefont {A.}~\bibnamefont
  {Butler}}, \bibinfo {author} {\bibfnamefont {D.~J.}\ \bibnamefont {Spence}},
  \ and\ \bibinfo {author} {\bibfnamefont {S.~M.}\ \bibnamefont {Hooker}},\
  }\href@noop {} {\bibfield  {journal} {\bibinfo  {journal} {Physical Review
  Letters}\ }\textbf {\bibinfo {volume} {89}},\ \bibinfo {pages} {185003}
  (\bibinfo {year} {2002})}\BibitemShut {NoStop}%
\bibitem [{\citenamefont {Leemans}\ \emph {et~al.}(2006)\citenamefont
  {Leemans}, \citenamefont {Nagler}, \citenamefont {Gonsalves}, \citenamefont
  {T~o th}, \citenamefont {Nakamura}, \citenamefont {Geddes}, \citenamefont
  {Esarey}, \citenamefont {Schroeder},\ and\ \citenamefont
  {Hooker}}]{Leemans:2006ux}%
  \BibitemOpen
  \bibfield  {author} {\bibinfo {author} {\bibfnamefont {W.~P.}\ \bibnamefont
  {Leemans}}, \bibinfo {author} {\bibfnamefont {B.}~\bibnamefont {Nagler}},
  \bibinfo {author} {\bibfnamefont {A.~J.}\ \bibnamefont {Gonsalves}}, \bibinfo
  {author} {\bibfnamefont {C.}~\bibnamefont {T~o th}}, \bibinfo {author}
  {\bibfnamefont {K.}~\bibnamefont {Nakamura}}, \bibinfo {author}
  {\bibfnamefont {C.~G.~R.}\ \bibnamefont {Geddes}}, \bibinfo {author}
  {\bibfnamefont {E.}~\bibnamefont {Esarey}}, \bibinfo {author} {\bibfnamefont
  {C.~B.}\ \bibnamefont {Schroeder}}, \ and\ \bibinfo {author} {\bibfnamefont
  {S.~M.}\ \bibnamefont {Hooker}},\ }\href@noop {} {\bibfield  {journal}
  {\bibinfo  {journal} {Nature Physics}\ }\textbf {\bibinfo {volume} {2}},\
  \bibinfo {pages} {696} (\bibinfo {year} {2006})}\BibitemShut {NoStop}%
\bibitem [{\citenamefont {Leemans}\ \emph {et~al.}(2014)\citenamefont
  {Leemans}, \citenamefont {Gonsalves}, \citenamefont {Mao}, \citenamefont
  {Nakamura}, \citenamefont {Benedetti}, \citenamefont {Schroeder},
  \citenamefont {Toth}, \citenamefont {Daniels}, \citenamefont {Mittelberger},
  \citenamefont {Bulanov}, \citenamefont {Vay}, \citenamefont {Geddes},\ and\
  \citenamefont {Esarey}}]{Leemans:2014kp}%
  \BibitemOpen
  \bibfield  {author} {\bibinfo {author} {\bibfnamefont {W.~P.}\ \bibnamefont
  {Leemans}}, \bibinfo {author} {\bibfnamefont {A.~J.}\ \bibnamefont
  {Gonsalves}}, \bibinfo {author} {\bibfnamefont {H.~S.}\ \bibnamefont {Mao}},
  \bibinfo {author} {\bibfnamefont {K.}~\bibnamefont {Nakamura}}, \bibinfo
  {author} {\bibfnamefont {C.}~\bibnamefont {Benedetti}}, \bibinfo {author}
  {\bibfnamefont {C.~B.}\ \bibnamefont {Schroeder}}, \bibinfo {author}
  {\bibfnamefont {C.}~\bibnamefont {Toth}}, \bibinfo {author} {\bibfnamefont
  {J.}~\bibnamefont {Daniels}}, \bibinfo {author} {\bibfnamefont {D.~E.}\
  \bibnamefont {Mittelberger}}, \bibinfo {author} {\bibfnamefont {S.~S.}\
  \bibnamefont {Bulanov}}, \bibinfo {author} {\bibfnamefont {J.~L.}\
  \bibnamefont {Vay}}, \bibinfo {author} {\bibfnamefont {C.~G.~R.}\
  \bibnamefont {Geddes}}, \ and\ \bibinfo {author} {\bibfnamefont
  {E.}~\bibnamefont {Esarey}},\ }\href@noop {} {\bibfield  {journal} {\bibinfo
  {journal} {Physical Review Letters}\ }\textbf {\bibinfo {volume} {113}},\
  \bibinfo {pages} {245002} (\bibinfo {year} {2014})}\BibitemShut {NoStop}%
\bibitem [{\citenamefont {Ju}\ \emph {et~al.}(2012)\citenamefont {Ju},
  \citenamefont {Svensson}, \citenamefont {D{\"o}pp}, \citenamefont {Ferrari},
  \citenamefont {Cassou}, \citenamefont {Neveu}, \citenamefont {Genoud},
  \citenamefont {Wojda}, \citenamefont {Burza}, \citenamefont {Persson},
  \citenamefont {Lundh}, \citenamefont {Wahlstrom},\ and\ \citenamefont
  {Cros}}]{Ju:2012jz}%
  \BibitemOpen
  \bibfield  {author} {\bibinfo {author} {\bibfnamefont {J.}~\bibnamefont
  {Ju}}, \bibinfo {author} {\bibfnamefont {K.}~\bibnamefont {Svensson}},
  \bibinfo {author} {\bibfnamefont {A.}~\bibnamefont {D{\"o}pp}}, \bibinfo
  {author} {\bibfnamefont {H.~E.}\ \bibnamefont {Ferrari}}, \bibinfo {author}
  {\bibfnamefont {K.}~\bibnamefont {Cassou}}, \bibinfo {author} {\bibfnamefont
  {O.}~\bibnamefont {Neveu}}, \bibinfo {author} {\bibfnamefont
  {G.}~\bibnamefont {Genoud}}, \bibinfo {author} {\bibfnamefont
  {F.}~\bibnamefont {Wojda}}, \bibinfo {author} {\bibfnamefont
  {M.}~\bibnamefont {Burza}}, \bibinfo {author} {\bibfnamefont
  {A.}~\bibnamefont {Persson}}, \bibinfo {author} {\bibfnamefont
  {O.}~\bibnamefont {Lundh}}, \bibinfo {author} {\bibfnamefont {C.~G.}\
  \bibnamefont {Wahlstrom}}, \ and\ \bibinfo {author} {\bibfnamefont
  {B.}~\bibnamefont {Cros}},\ }\href@noop {} {\bibfield  {journal} {\bibinfo
  {journal} {Applied Physics Letters}\ }\textbf {\bibinfo {volume} {100}},\
  \bibinfo {pages} {191106} (\bibinfo {year} {2012})}\BibitemShut {NoStop}%
\bibitem [{\citenamefont {Corde}\ \emph {et~al.}(2013)\citenamefont {Corde},
  \citenamefont {Lifschitz}, \citenamefont {Lambert}, \citenamefont {Phuoc},
  \citenamefont {Davoine}, \citenamefont {Lehe}, \citenamefont {Douillet},
  \citenamefont {Rousse}, \citenamefont {Malka},\ and\ \citenamefont
  {Thaury}}]{Corde:2013gj}%
  \BibitemOpen
  \bibfield  {author} {\bibinfo {author} {\bibfnamefont {S.}~\bibnamefont
  {Corde}}, \bibinfo {author} {\bibfnamefont {A.}~\bibnamefont {Lifschitz}},
  \bibinfo {author} {\bibfnamefont {G.}~\bibnamefont {Lambert}}, \bibinfo
  {author} {\bibfnamefont {K.~T.}\ \bibnamefont {Phuoc}}, \bibinfo {author}
  {\bibfnamefont {X.}~\bibnamefont {Davoine}}, \bibinfo {author} {\bibfnamefont
  {R.}~\bibnamefont {Lehe}}, \bibinfo {author} {\bibfnamefont {D.}~\bibnamefont
  {Douillet}}, \bibinfo {author} {\bibfnamefont {A.}~\bibnamefont {Rousse}},
  \bibinfo {author} {\bibfnamefont {V.}~\bibnamefont {Malka}}, \ and\ \bibinfo
  {author} {\bibfnamefont {C.}~\bibnamefont {Thaury}},\ }\href@noop {}
  {\bibfield  {journal} {\bibinfo  {journal} {Nature Communications}\ }\textbf
  {\bibinfo {volume} {4}},\ \bibinfo {pages} {1501} (\bibinfo {year}
  {2013})}\BibitemShut {NoStop}%
\bibitem [{\citenamefont {Ju}\ and\ \citenamefont {Cros}(2012)}]{Ju:2012es}%
  \BibitemOpen
  \bibfield  {author} {\bibinfo {author} {\bibfnamefont {J.}~\bibnamefont
  {Ju}}\ and\ \bibinfo {author} {\bibfnamefont {B.}~\bibnamefont {Cros}},\
  }\href@noop {} {\bibfield  {journal} {\bibinfo  {journal} {Journal of Applied
  Physics}\ }\textbf {\bibinfo {volume} {112}},\ \bibinfo {pages} {113102}
  (\bibinfo {year} {2012})}\BibitemShut {NoStop}%
\bibitem [{\citenamefont {Semushin}\ and\ \citenamefont
  {Malka}(2001)}]{Semushin:2001fl}%
  \BibitemOpen
  \bibfield  {author} {\bibinfo {author} {\bibfnamefont {S.}~\bibnamefont
  {Semushin}}\ and\ \bibinfo {author} {\bibfnamefont {V.}~\bibnamefont
  {Malka}},\ }\href@noop {} {\bibfield  {journal} {\bibinfo  {journal} {Review
  of Scientific Instruments}\ }\textbf {\bibinfo {volume} {72}},\ \bibinfo
  {pages} {2961} (\bibinfo {year} {2001})}\BibitemShut {NoStop}%
\bibitem [{\citenamefont {Schmid}\ and\ \citenamefont
  {Veisz}(2012)}]{Schmid:2012ki}%
  \BibitemOpen
  \bibfield  {author} {\bibinfo {author} {\bibfnamefont {K.}~\bibnamefont
  {Schmid}}\ and\ \bibinfo {author} {\bibfnamefont {L.}~\bibnamefont {Veisz}},\
  }\href@noop {} {\bibfield  {journal} {\bibinfo  {journal} {Review of
  Scientific Instruments}\ }\textbf {\bibinfo {volume} {83}},\ \bibinfo {pages}
  {053304} (\bibinfo {year} {2012})}\BibitemShut {NoStop}%
\bibitem [{\citenamefont {Najmudin}\ \emph {et~al.}(2014)\citenamefont
  {Najmudin}, \citenamefont {Kneip}, \citenamefont {Bloom}, \citenamefont
  {Mangles}, \citenamefont {Chekhlov}, \citenamefont {Dangor}, \citenamefont
  {D{\"o}pp}, \citenamefont {Ertel}, \citenamefont {Hawkes}, \citenamefont
  {Holloway}, \citenamefont {Hooker}, \citenamefont {Jiang}, \citenamefont
  {Lopes}, \citenamefont {Nakamura}, \citenamefont {Norreys}, \citenamefont
  {Rajeev}, \citenamefont {Russo}, \citenamefont {Streeter}, \citenamefont
  {Symes},\ and\ \citenamefont {Wing}}]{Najmudin:2014uu}%
  \BibitemOpen
  \bibfield  {author} {\bibinfo {author} {\bibfnamefont {Z.}~\bibnamefont
  {Najmudin}}, \bibinfo {author} {\bibfnamefont {S.}~\bibnamefont {Kneip}},
  \bibinfo {author} {\bibfnamefont {M.~S.}\ \bibnamefont {Bloom}}, \bibinfo
  {author} {\bibfnamefont {S.~P.~D.}\ \bibnamefont {Mangles}}, \bibinfo
  {author} {\bibfnamefont {O.}~\bibnamefont {Chekhlov}}, \bibinfo {author}
  {\bibfnamefont {A.~E.}\ \bibnamefont {Dangor}}, \bibinfo {author}
  {\bibfnamefont {A.}~\bibnamefont {D{\"o}pp}}, \bibinfo {author}
  {\bibfnamefont {K.}~\bibnamefont {Ertel}}, \bibinfo {author} {\bibfnamefont
  {S.~J.}\ \bibnamefont {Hawkes}}, \bibinfo {author} {\bibfnamefont
  {J.}~\bibnamefont {Holloway}}, \bibinfo {author} {\bibfnamefont {C.~J.}\
  \bibnamefont {Hooker}}, \bibinfo {author} {\bibfnamefont {J.}~\bibnamefont
  {Jiang}}, \bibinfo {author} {\bibfnamefont {N.~C.}\ \bibnamefont {Lopes}},
  \bibinfo {author} {\bibfnamefont {H.}~\bibnamefont {Nakamura}}, \bibinfo
  {author} {\bibfnamefont {P.~A.}\ \bibnamefont {Norreys}}, \bibinfo {author}
  {\bibfnamefont {P.~P.}\ \bibnamefont {Rajeev}}, \bibinfo {author}
  {\bibfnamefont {C.}~\bibnamefont {Russo}}, \bibinfo {author} {\bibfnamefont
  {M.~J.~V.}\ \bibnamefont {Streeter}}, \bibinfo {author} {\bibfnamefont
  {D.~R.}\ \bibnamefont {Symes}}, \ and\ \bibinfo {author} {\bibfnamefont
  {M.}~\bibnamefont {Wing}},\ }\href@noop {} {\bibfield  {journal} {\bibinfo
  {journal} {Philosophical Transactions of the Royal Society A: Mathematical,
  Physical and Engineering Sciences}\ }\textbf {\bibinfo {volume} {372}}
  (\bibinfo {year} {2014})}\BibitemShut {NoStop}%
\bibitem [{\citenamefont {Sylla}\ \emph {et~al.}(2012)\citenamefont {Sylla},
  \citenamefont {Veltcheva}, \citenamefont {Kahaly}, \citenamefont {Flacco},\
  and\ \citenamefont {Malka}}]{Sylla:2012hl}%
  \BibitemOpen
  \bibfield  {author} {\bibinfo {author} {\bibfnamefont {F.}~\bibnamefont
  {Sylla}}, \bibinfo {author} {\bibfnamefont {M.}~\bibnamefont {Veltcheva}},
  \bibinfo {author} {\bibfnamefont {S.}~\bibnamefont {Kahaly}}, \bibinfo
  {author} {\bibfnamefont {A.}~\bibnamefont {Flacco}}, \ and\ \bibinfo {author}
  {\bibfnamefont {V.}~\bibnamefont {Malka}},\ }\href@noop {} {\bibfield
  {journal} {\bibinfo  {journal} {Review of Scientific Instruments}\ }\textbf
  {\bibinfo {volume} {83}},\ \bibinfo {pages} {033507} (\bibinfo {year}
  {2012})}\BibitemShut {NoStop}%
\bibitem [{\citenamefont {Golovin}\ \emph {et~al.}(2015)\citenamefont
  {Golovin}, \citenamefont {Chen}, \citenamefont {Powers}, \citenamefont {Liu},
  \citenamefont {Banerjee}, \citenamefont {Zhang}, \citenamefont {Zeng},
  \citenamefont {Sheng},\ and\ \citenamefont {Umstadter}}]{Golovin:2015hc}%
  \BibitemOpen
  \bibfield  {author} {\bibinfo {author} {\bibfnamefont {G.}~\bibnamefont
  {Golovin}}, \bibinfo {author} {\bibfnamefont {S.}~\bibnamefont {Chen}},
  \bibinfo {author} {\bibfnamefont {N.}~\bibnamefont {Powers}}, \bibinfo
  {author} {\bibfnamefont {C.}~\bibnamefont {Liu}}, \bibinfo {author}
  {\bibfnamefont {S.}~\bibnamefont {Banerjee}}, \bibinfo {author}
  {\bibfnamefont {J.}~\bibnamefont {Zhang}}, \bibinfo {author} {\bibfnamefont
  {M.}~\bibnamefont {Zeng}}, \bibinfo {author} {\bibfnamefont {Z.}~\bibnamefont
  {Sheng}}, \ and\ \bibinfo {author} {\bibfnamefont {D.}~\bibnamefont
  {Umstadter}},\ }\href@noop {} {\bibfield  {journal} {\bibinfo  {journal}
  {Physical Review Special Topics-Accelerators and Beams}\ }\textbf {\bibinfo
  {volume} {18}},\ \bibinfo {pages} {011301} (\bibinfo {year}
  {2015})}\BibitemShut {NoStop}%
\bibitem [{\citenamefont {Sachs}(1992)}]{Sachs:1992cb}%
  \BibitemOpen
  \bibfield  {author} {\bibinfo {author} {\bibfnamefont {E.}~\bibnamefont
  {Sachs}},\ }\href@noop {} {\bibfield  {journal} {\bibinfo  {journal} {Journal
  of Engineering for Industry}\ ,\ \bibinfo {pages} {201}} (\bibinfo {year}
  {1992})}\BibitemShut {NoStop}%
\bibitem [{\citenamefont {Levy}\ \emph {et~al.}(2003)\citenamefont {Levy},
  \citenamefont {Schindel},\ and\ \citenamefont {Kruth}}]{Levy:2003ie}%
  \BibitemOpen
  \bibfield  {author} {\bibinfo {author} {\bibfnamefont {G.~N.}\ \bibnamefont
  {Levy}}, \bibinfo {author} {\bibfnamefont {R.}~\bibnamefont {Schindel}}, \
  and\ \bibinfo {author} {\bibfnamefont {J.~P.}\ \bibnamefont {Kruth}},\
  }\href@noop {} {\bibfield  {journal} {\bibinfo  {journal} {CIRP Annals -
  Manufacturing Technology}\ }\textbf {\bibinfo {volume} {52}},\ \bibinfo
  {pages} {589} (\bibinfo {year} {2003})}\BibitemShut {NoStop}%
\bibitem [{\citenamefont {Wong}\ and\ \citenamefont
  {Hernandez}(2012)}]{Wong:2012iq}%
  \BibitemOpen
  \bibfield  {author} {\bibinfo {author} {\bibfnamefont {K.~V.}\ \bibnamefont
  {Wong}}\ and\ \bibinfo {author} {\bibfnamefont {A.}~\bibnamefont
  {Hernandez}},\ }\href@noop {} {\bibfield  {journal} {\bibinfo  {journal}
  {ISRN Mechanical Engineering}\ }\textbf {\bibinfo {volume} {2012}},\ \bibinfo
  {pages} {1} (\bibinfo {year} {2012})}\BibitemShut {NoStop}%
\bibitem [{\citenamefont {Conner}\ \emph {et~al.}(2014)\citenamefont {Conner},
  \citenamefont {Manogharan}, \citenamefont {Martof}, \citenamefont {Rodomsky},
  \citenamefont {Rodomsky}, \citenamefont {Jordan},\ and\ \citenamefont
  {Limperos}}]{Conner:2014hr}%
  \BibitemOpen
  \bibfield  {author} {\bibinfo {author} {\bibfnamefont {B.~P.}\ \bibnamefont
  {Conner}}, \bibinfo {author} {\bibfnamefont {G.~P.}\ \bibnamefont
  {Manogharan}}, \bibinfo {author} {\bibfnamefont {A.~N.}\ \bibnamefont
  {Martof}}, \bibinfo {author} {\bibfnamefont {L.~M.}\ \bibnamefont
  {Rodomsky}}, \bibinfo {author} {\bibfnamefont {C.~M.}\ \bibnamefont
  {Rodomsky}}, \bibinfo {author} {\bibfnamefont {D.~C.}\ \bibnamefont
  {Jordan}}, \ and\ \bibinfo {author} {\bibfnamefont {J.~W.}\ \bibnamefont
  {Limperos}},\ }\href@noop {} {\bibfield  {journal} {\bibinfo  {journal}
  {Additive Manufacturing}\ ,\ \bibinfo {pages} {1}} (\bibinfo {year}
  {2014})}\BibitemShut {NoStop}%
\bibitem [{\citenamefont {Berman}(2014)}]{Berman:2014ko}%
  \BibitemOpen
  \bibfield  {author} {\bibinfo {author} {\bibfnamefont {B.}~\bibnamefont
  {Berman}},\ }\href@noop {} {\bibfield  {journal} {\bibinfo  {journal}
  {Business Horizons}\ }\textbf {\bibinfo {volume} {55}},\ \bibinfo {pages}
  {155} (\bibinfo {year} {2014})}\BibitemShut {NoStop}%
\bibitem [{\citenamefont {Hornick}\ and\ \citenamefont
  {Roland}(2013)}]{Hornick:2013wx}%
  \BibitemOpen
  \bibfield  {author} {\bibinfo {author} {\bibfnamefont {J.}~\bibnamefont
  {Hornick}}\ and\ \bibinfo {author} {\bibfnamefont {D.}~\bibnamefont
  {Roland}},\ }\href@noop {} {\bibfield  {journal} {\bibinfo  {journal}
  {3dprintingindustry.com}\ } (\bibinfo {year} {2013})}\BibitemShut {NoStop}%
\bibitem [{\citenamefont {Gross}\ \emph {et~al.}(2014)\citenamefont {Gross},
  \citenamefont {Erkal}, \citenamefont {Lockwood}, \citenamefont {Chen},\ and\
  \citenamefont {Spence}}]{Gross:2014jxa}%
  \BibitemOpen
  \bibfield  {author} {\bibinfo {author} {\bibfnamefont {B.~C.}\ \bibnamefont
  {Gross}}, \bibinfo {author} {\bibfnamefont {J.~L.}\ \bibnamefont {Erkal}},
  \bibinfo {author} {\bibfnamefont {S.~Y.}\ \bibnamefont {Lockwood}}, \bibinfo
  {author} {\bibfnamefont {C.}~\bibnamefont {Chen}}, \ and\ \bibinfo {author}
  {\bibfnamefont {D.~M.}\ \bibnamefont {Spence}},\ }\href@noop {} {\bibfield
  {journal} {\bibinfo  {journal} {Analytical Chemistry}\ }\textbf {\bibinfo
  {volume} {86}},\ \bibinfo {pages} {3240} (\bibinfo {year}
  {2014})}\BibitemShut {NoStop}%
\bibitem [{\citenamefont {Gao}\ \emph {et~al.}(2015)\citenamefont {Gao},
  \citenamefont {Zhang}, \citenamefont {Ramanujan}, \citenamefont {Ramani},
  \citenamefont {Chen}, \citenamefont {Williams}, \citenamefont {Wang},
  \citenamefont {Shin}, \citenamefont {Zhang},\ and\ \citenamefont
  {Zavattieri}}]{Gao:2015bs}%
  \BibitemOpen
  \bibfield  {author} {\bibinfo {author} {\bibfnamefont {W.}~\bibnamefont
  {Gao}}, \bibinfo {author} {\bibfnamefont {Y.}~\bibnamefont {Zhang}}, \bibinfo
  {author} {\bibfnamefont {D.}~\bibnamefont {Ramanujan}}, \bibinfo {author}
  {\bibfnamefont {K.}~\bibnamefont {Ramani}}, \bibinfo {author} {\bibfnamefont
  {Y.}~\bibnamefont {Chen}}, \bibinfo {author} {\bibfnamefont {C.~B.}\
  \bibnamefont {Williams}}, \bibinfo {author} {\bibfnamefont {C.~C.~L.}\
  \bibnamefont {Wang}}, \bibinfo {author} {\bibfnamefont {Y.~C.}\ \bibnamefont
  {Shin}}, \bibinfo {author} {\bibfnamefont {S.}~\bibnamefont {Zhang}}, \ and\
  \bibinfo {author} {\bibfnamefont {P.~D.}\ \bibnamefont {Zavattieri}},\
  }\href@noop {} {\bibfield  {journal} {\bibinfo  {journal} {Computer-Aided
  Design}\ ,\ \bibinfo {pages} {1}} (\bibinfo {year} {2015})}\BibitemShut
  {NoStop}%
\bibitem [{\citenamefont {Crump}(1992)}]{Crump:1992ti}%
  \BibitemOpen
  \bibfield  {author} {\bibinfo {author} {\bibfnamefont {S.~S.}\ \bibnamefont
  {Crump}},\ }\href@noop {} {\ ,\ \bibinfo {pages} {1} (\bibinfo {year}
  {1992})}\BibitemShut {NoStop}%
\bibitem [{\citenamefont {Hull}(1986)}]{Hull:1986vi}%
  \BibitemOpen
  \bibfield  {author} {\bibinfo {author} {\bibfnamefont {C.~W.}\ \bibnamefont
  {Hull}},\ }\href@noop {} {\ ,\ \bibinfo {pages} {1} (\bibinfo {year}
  {1986})}\BibitemShut {NoStop}%
\bibitem [{\citenamefont {Jacobs}(1992)}]{Jacobs:1992wg}%
  \BibitemOpen
  \bibfield  {author} {\bibinfo {author} {\bibfnamefont {P.~F.}\ \bibnamefont
  {Jacobs}},\ }\href@noop {} {\emph {\bibinfo {title} {{Rapid prototyping {\&}
  manufacturing: fundamentals of stereolithography}}}}\ (\bibinfo  {publisher}
  {Society of Manufacturing Engineers},\ \bibinfo {year} {1992})\BibitemShut
  {NoStop}%
\bibitem [{\citenamefont {Deckard}(1989)}]{Deckard:1989wj}%
  \BibitemOpen
  \bibfield  {author} {\bibinfo {author} {\bibfnamefont {C.~R.}\ \bibnamefont
  {Deckard}},\ }\href@noop {} {\ ,\ \bibinfo {pages} {1} (\bibinfo {year}
  {1989})}\BibitemShut {NoStop}%
\bibitem [{\citenamefont {Kruth}\ \emph {et~al.}(2005)\citenamefont {Kruth},
  \citenamefont {Mercelis}, \citenamefont {Van~Vaerenbergh}, \citenamefont
  {Froyen},\ and\ \citenamefont {Rombouts}}]{Kruth:2005hv}%
  \BibitemOpen
  \bibfield  {author} {\bibinfo {author} {\bibfnamefont {J.~P.}\ \bibnamefont
  {Kruth}}, \bibinfo {author} {\bibfnamefont {P.}~\bibnamefont {Mercelis}},
  \bibinfo {author} {\bibfnamefont {J.}~\bibnamefont {Van~Vaerenbergh}},
  \bibinfo {author} {\bibfnamefont {L.}~\bibnamefont {Froyen}}, \ and\ \bibinfo
  {author} {\bibfnamefont {M.}~\bibnamefont {Rombouts}},\ }\href@noop {}
  {\bibfield  {journal} {\bibinfo  {journal} {Rapid Prototyping Journal}\
  }\textbf {\bibinfo {volume} {11}},\ \bibinfo {pages} {26} (\bibinfo {year}
  {2005})}\BibitemShut {NoStop}%
\bibitem [{\citenamefont {Vargas}\ \emph {et~al.}(2014)\citenamefont {Vargas},
  \citenamefont {Schumaker}, \citenamefont {He}, \citenamefont {Zhao},
  \citenamefont {Behm}, \citenamefont {Chvykov}, \citenamefont {Hou},
  \citenamefont {Krushelnick}, \citenamefont {Maksimchuk}, \citenamefont
  {Yanovsky},\ and\ \citenamefont {Thomas}}]{Vargas:2014dg}%
  \BibitemOpen
  \bibfield  {author} {\bibinfo {author} {\bibfnamefont {M.}~\bibnamefont
  {Vargas}}, \bibinfo {author} {\bibfnamefont {W.}~\bibnamefont {Schumaker}},
  \bibinfo {author} {\bibfnamefont {Z.~H.}\ \bibnamefont {He}}, \bibinfo
  {author} {\bibfnamefont {Z.}~\bibnamefont {Zhao}}, \bibinfo {author}
  {\bibfnamefont {K.}~\bibnamefont {Behm}}, \bibinfo {author} {\bibfnamefont
  {V.}~\bibnamefont {Chvykov}}, \bibinfo {author} {\bibfnamefont
  {B.}~\bibnamefont {Hou}}, \bibinfo {author} {\bibfnamefont {K.}~\bibnamefont
  {Krushelnick}}, \bibinfo {author} {\bibfnamefont {A.}~\bibnamefont
  {Maksimchuk}}, \bibinfo {author} {\bibfnamefont {V.}~\bibnamefont
  {Yanovsky}}, \ and\ \bibinfo {author} {\bibfnamefont {A.~G.~R.}\ \bibnamefont
  {Thomas}},\ }\href@noop {} {\bibfield  {journal} {\bibinfo  {journal}
  {Applied Physics Letters}\ }\textbf {\bibinfo {volume} {104}},\ \bibinfo
  {pages} {174103} (\bibinfo {year} {2014})}\BibitemShut {NoStop}%
\bibitem [{\citenamefont {Jolly}\ \emph {et~al.}(2012)\citenamefont {Jolly},
  \citenamefont {He}, \citenamefont {McGuffey}, \citenamefont {Schumaker},
  \citenamefont {Krushelnick},\ and\ \citenamefont {Thomas}}]{Jolly:2012ep}%
  \BibitemOpen
  \bibfield  {author} {\bibinfo {author} {\bibfnamefont {S.~W.}\ \bibnamefont
  {Jolly}}, \bibinfo {author} {\bibfnamefont {Z.}~\bibnamefont {He}}, \bibinfo
  {author} {\bibfnamefont {C.}~\bibnamefont {McGuffey}}, \bibinfo {author}
  {\bibfnamefont {W.}~\bibnamefont {Schumaker}}, \bibinfo {author}
  {\bibfnamefont {K.}~\bibnamefont {Krushelnick}}, \ and\ \bibinfo {author}
  {\bibfnamefont {A.~G.~R.}\ \bibnamefont {Thomas}},\ }\href@noop {} {\bibfield
   {journal} {\bibinfo  {journal} {Review of Scientific Instruments}\ }\textbf
  {\bibinfo {volume} {83}},\ \bibinfo {pages} {073503} (\bibinfo {year}
  {2012})}\BibitemShut {NoStop}%
\bibitem [{\citenamefont {Hernandez}(2015)}]{Hernandez:2015wc}%
  \BibitemOpen
  \bibfield  {author} {\bibinfo {author} {\bibfnamefont {D.~D.}\ \bibnamefont
  {Hernandez}},\ }\href@noop {} {\bibfield  {journal} {\bibinfo  {journal}
  {International Journal of Aviation, Aeronautics and Aerospace}\ }\textbf
  {\bibinfo {volume} {2}},\ \bibinfo {pages} {1} (\bibinfo {year}
  {2015})}\BibitemShut {NoStop}%
\bibitem [{\citenamefont {Benattar}\ \emph {et~al.}(1979)\citenamefont
  {Benattar}, \citenamefont {Popovics},\ and\ \citenamefont
  {Sigel}}]{Benattar:1979uha}%
  \BibitemOpen
  \bibfield  {author} {\bibinfo {author} {\bibfnamefont {R.}~\bibnamefont
  {Benattar}}, \bibinfo {author} {\bibfnamefont {C.}~\bibnamefont {Popovics}},
  \ and\ \bibinfo {author} {\bibfnamefont {R.}~\bibnamefont {Sigel}},\
  }\href@noop {} {\bibfield  {journal} {\bibinfo  {journal} {Review of
  Scientific Instruments}\ }\textbf {\bibinfo {volume} {50}},\ \bibinfo {pages}
  {1583} (\bibinfo {year} {1979})}\BibitemShut {NoStop}%
\bibitem [{\citenamefont {Povilus}\ \emph {et~al.}(2014)\citenamefont
  {Povilus}, \citenamefont {Wurden}, \citenamefont {Vendeiro}, \citenamefont
  {Baquero-Ruiz},\ and\ \citenamefont {Fajans}}]{Povilus:2014jz}%
  \BibitemOpen
  \bibfield  {author} {\bibinfo {author} {\bibfnamefont {A.~P.}\ \bibnamefont
  {Povilus}}, \bibinfo {author} {\bibfnamefont {C.~J.}\ \bibnamefont {Wurden}},
  \bibinfo {author} {\bibfnamefont {Z.}~\bibnamefont {Vendeiro}}, \bibinfo
  {author} {\bibfnamefont {M.}~\bibnamefont {Baquero-Ruiz}}, \ and\ \bibinfo
  {author} {\bibfnamefont {J.}~\bibnamefont {Fajans}},\ }\href@noop {}
  {\bibfield  {journal} {\bibinfo  {journal} {Journal of Vacuum Science {\&}
  Technology A: Vacuum, Surfaces, and Films}\ }\textbf {\bibinfo {volume}
  {32}},\ \bibinfo {pages} {033001} (\bibinfo {year} {2014})}\BibitemShut
  {NoStop}%
\end{thebibliography}

%

\end{document}